\documentclass[reprint,floatfix,aps,prd,showpacs,footinbib,amsmath,amssymb,amsfonts,superscriptaddress]{revtex4-1}
\usepackage{bm}
\usepackage{graphicx,color}
\usepackage{physics}
\usepackage{dsfont}
\usepackage[normalem]{ulem}
\usepackage{hyperref,url}
\usepackage{mathrsfs}
\usepackage{calrsfs}
\renewcommand{\vec}[1]{\mathbf{#1}}
\usepackage{varioref}
\usepackage[active]{srcltx}

\expandafter\ifx\csname package@font\endcsname\relax\else
\expandafter\expandafter
\expandafter\usepackage
\expandafter\expandafter
\expandafter{\csname package@font\endcsname}%
\fi
\hypersetup{
	colorlinks=true,       % false: boxed links; true: colored links
	linkcolor=blue,          % color of internal links
	citecolor=blue,        % color of links to bibliography
}

\usepackage[section]{placeins}

\usepackage{fancybox}
\labelformat{figure}{Fig.~#1} 

\DeclareMathOperator\ei{Ei}
\begin{document}
\title{Divergence of entanglement entropy in quantum systems: Zero-modes}
\author{S. Mahesh Chandran} 
\email{maheshchandran14895@gmail.com}
\affiliation{Department of Physics, Indian Institute of Technology Bombay, Mumbai 400076, India}
\affiliation{School of Physics, Indian Institute of Science Education and Research Thiruvananthapuram, Trivandrum 695551, India}
\author{S. Shankaranarayanan}
\email{shanki@phy.iitb.ac.in}
\affiliation{Department of Physics, Indian Institute of Technology Bombay, Mumbai 400076, India}
%%%
\begin{abstract}
To this day, von Neumann definition of entropy remains the most popular measure of quantum entanglement. Much of the literature on entanglement entropy, particularly in the context of field theory, has focused on isolating the UV divergences. Zero-mode divergences of the entanglement entropy are less studied in this context, and apart from being easier to isolate, they offer an interesting insight into the physics of the system. To gain a better understanding of the system in this limit, we develop the free particle approximation of Harmonic oscillator, with which we investigate the properties of entropy divergence in continuous bi-partite quantum systems such as the coupled Harmonic oscillators and the Hydrogen atom. We also show zero-mode divergence of the entropy of environment-induced entanglement in a tri-partite oscillator system. We discuss the implications of our result for field theory and IR structure of gravity.
\end{abstract}

\maketitle

\section{Introduction}

Entanglement is the name given to purely quantum correlations between systems or particles. These quantum correlations are independent of any representation and the spatial distance between the sub-systems~\cite{RevModPhys.81.865,RevModPhys.82.277,1751-8121-42-50-504005}. In other words, two sub-systems (or particles)  could be parted by light years in distance and measurement of one sub-system can provide complete information about the other system. Recently, quantum entanglement has been verified over a thousand kilometres for a 2-photon pair~\cite{2016-Yin.etal-Science}. This raises two crucial questions:  How does entanglement store information about the possible experimental outcome without having information about the individual sub-systems? Is entanglement oblivious to space and time? 

The current lack of a fundamental understanding of these questions prevents us from addressing questions related to entanglement in field theory. In general, entanglement is an essential feature of quantum field theory~\cite{2017-Unruh.Wald-RPP}. More importantly, the divergence of a two-point function of a scalar field implies that the fields are highly entangled, since:
\begin{equation}
\langle \Psi | \hat{\phi}(x_1) \hat{\phi}(x_2) | \Psi \rangle 
\neq \langle \Psi | \hat{\phi}(x_1) | \Psi \rangle \langle \Psi | \hat{\phi}(x_2) | \Psi \rangle
\end{equation}
In other words, irrespective of the space-time or quantum state, entanglement always exists between the quantum field. Entanglement entropy $(S_{\rm ent}$) of a scalar field, defined as the entropy of the reduced state of a subregion, scales as the boundary area ${\cal A}$ of the subregion \cite{1986-Bombelli.etal-PRD,*PhysRevLett.71.666,*2006-Calabrese.Cardy-IJQI,*bhole,*Solodukhin2011}.  The area-law appears to be valid for all 1-dimensional gapped many-body systems  and some gapless systems in higher dimensions~\cite{RevModPhys.81.865,RevModPhys.82.277,2008-Wolf.etal-PRL,*2012-Peschel-BJP}. The scaling of the entanglement entropy reveals most of the long-distance properties of the system~\cite{RevModPhys.69.315,*2003-Vojta-RPP,*2003-Vidal-PRL,*2008-Amico.etal-RMP,*PhysRevLett.93.250404}.

Irrespective of the physical (black-holes or condensed matter) system, the prefactor in the area-law depends on the UV cut-off of the theory~\cite{1996-Larsen.Wilczek-NPB,*PhysRevD.82.124025,*Nesterov2010,*NESTEROV2011141}. Thus, the value of $S_{ent}/{\cal A}$ is set by the cut-off and increases when the cut-off is increased. Physically, the increase is due to entanglement between more degrees of freedom. However, if the high-energy modes decouple from low-energy phenomena, then it is unsettling as to why quantifying entanglement for a macroscopic system should strongly depend on the UV cut-off.  

Recent studies have laid bare the inevitable contribution of zero-modes in this context~\cite{PhysRevD.90.044058,Yazdi2017,*Speranza2016,*1751-8121-42-50-504007,*PhysRevD.92.105022}, thereby prompting extensive analysis of the behaviour of the entanglement entropy in the infra-red. 
%The peculiarity of IR divergences is that they are rarely observed or studied %in physical systems because zero-modes are practically difficult to achieve.
The peculiarity of these divergences is that the zero-modes are independent of the spatial coordinates~\cite{PhysRevD.39.3642}.
The independence of the spatial coordinates prevents us from obtaining some crucial information about the system which could otherwise help explain the underlying physics behind such divergences, and how to tame these divergences. For instance, the zero modes (massless states) lead to the IR divergences of the $\mathcal{S}$-matrix in QED~\cite{PhysRevD.96.085002}. On the other hand, UV-divergence implies that the physics of the system is highly sensitive to the UV-cutoff employed, and this causes discrepancies, especially at high energies. 

In general, it is an arduous task to obtain an analytic expression for entanglement entropy except for a few special cases like $(1 + 1)-$dimensional CFTs \cite{1751-8121-42-50-504005}. To get a better
analytical control and understanding of zero-mode divergence, in this work, we 
consider the ubiquitous Coulomb potential and coupled Harmonic oscillators. The Coulomb potential is non-linear, and in the many-particle case, it is the two-body potential. This potential can be obtained from reducing the N-body problem rigorously to a combinatorial number of two-body interactions.  In QFT, a field can be treated as an infinite number of coupled harmonic oscillators. We should in principle be able to extend the analysis on coupled harmonic oscillators directly to quantum fields. As mentioned above --- to weigh in on the nature of divergence of entropy in physical systems --- we consider Hydrogen atom and coupled harmonic oscillators, whose entanglement entropy can be analytically calculated.

In Section \hyperref[2]{II}, we investigate the properties of entanglement in a system of two harmonic oscillators that have a negative coupling constant. The resultant entropy is found to diverge in the free particle limit (or the zero frequency mode) of the system in normal co-ordinates. To further investigate the origin of these divergences, we need to see how the entropy behaves as the system develops zero-modes. To this end, in Section \hyperref[3]{III}, we obtain the free particle approximation of harmonic oscillator which we then utilize in Section \hyperref[4]{IV} to provide the zero-mode analysis of the coupled harmonic oscillator system. In Section \hyperref[5]{V}, we use this approximation to study the Hydrogen atom in a new light, and further confirm that the entropy for such physical systems diverges due to zero-modes, as opposed to the commonly held view of UV-divergence of entropy. 

Furthermore, we investigate the nature of divergence of entanglement entropy when the system in question is canonically transformed to a new Hamiltonian. In Section \hyperref[6]{VI}, we map the Hydrogen atom to a harmonic oscillator using a well-known transformation\cite{doi:10.1119/1.17065}, and see if the properties of entanglement entropy are preserved. In Section \hyperref[7]{VII}, we continue to probe for zero-mode divergence of entanglement entropy in a system where entanglement between two uncoupled oscillators is mediated by an oscillator sink, as opposed to the cases of the Hydrogen atom and coupled Harmonic oscillator where the sub-systems are directly coupled. In Section \hyperref[8]{VIII}, we conclude by briefly discussing the importance of these results. 
%, all in an attempt to provide a more coherent commentary on the divergence of entanglement entropy in quantum systems.

\section{Coupled Harmonic Oscillator : A Quick Review}
\label{2}

As a warm-up and to make sure that the definitions are transparent, we calculate the entanglement entropy for coupled harmonic oscillators with a negative coupling constant~\cite{2005-Plenio.etal-PRL}. Note that the coupled harmonic oscillator with positive coupling constant is well studied (see, for instance, \cite{PhysRevLett.71.666}).  The Hamiltonian  is given by:
{\small
\begin{equation}
    H=-\frac{\hbar^2}{2m}\bigg[\frac{\partial^2}{\partial x_1^2}+\frac{\partial^2}{\partial x_2^2}\bigg]+\frac{1}{2}m\bigg[\omega_0^2(x_1^2+x_2^2) - \omega_1^2(x_1-x_2)^2\bigg] 
\end{equation}
}
\!\!\! where $\omega_0^2$ and $\omega_1^2$ are positive constants. The normalized ground state wave-functions in the  normal mode coordinates $x_\pm=(x_1\pm x_2)/\sqrt{2}$ is:
\begin{equation}
    \Psi_0(x_+,x_-)=\frac{(\beta_+\beta_-)^{1/4}}{\sqrt{\pi}}\exp{-\frac{\beta_+x_+^2}{2}-\frac{\beta_-x_-^2}{2}},
\end{equation}
where
\begin{equation}
 \beta_\pm= \frac{m\omega_\pm}{\hbar}; \qquad  \omega_+=\omega_0; \qquad  \omega_- =\sqrt{\omega_0^2-2\omega_1^2}
\end{equation}
In the original coordinates $(x_1, x_2)$, the above ground state is entangled, i. e.,
\begin{multline}
     \Psi_0(x_1,x_2)=\frac{(\beta_+\beta_-)^{1/4}}{\sqrt{\pi}}\exp{-\frac{\beta_+(x_1+x_2)^2}{4}}\\\times\exp{-\frac{\beta_-(x_1-x_2)^2}{4}}
\end{multline}
The density matrix $\rho=\ket{\Psi}\bra{\Psi}$ is symmetric, and we can simply proceed to trace out any one of the sub-systems to obtain the reduced density matrix of the other:
\begin{align}
    \rho_1(x_1,x_1')&=\int\limits_{-\infty}^{\infty}dx_2\Psi_0^*(x_1',x_2)\Psi_0(x_1,x_2) \\
    &=\sqrt{\frac{\gamma_1-\gamma_2}{\pi}}\exp{-\gamma_1\frac{(x_1^2+x_1'^2)}{2}-\gamma_2x_1x_1'} \, , \nonumber
\end{align}
where $\gamma_1$ and $\gamma_2$ are given by:
\begin{align}
    \gamma_1&=\frac{\beta_+^2+\beta_-^2+6\beta_+\beta_-}{4(\beta_++\beta_-)}\\
    \gamma_2&=\frac{(\beta_+-\beta_-)^2}{4(\beta_++\beta_-)}
\end{align}
In order to find the eigenvalues of $\rho_1(x_1,x_1')$, we must solve the following integral equation for $p_n$:
\begin{equation}
    \int\limits_{-\infty}^{\infty}dx_1'\rho_1(x_1,x_1')f_n(x_1')=p_nf_n(x_1)
\end{equation}
The solution for the above integral equation is:
\begin{align}
    p_n&=(1-\xi)\xi^n,\\
    f_n(x)&=H_n(\sqrt{\varrho}x)\exp{-\varrho\frac{x^2}{2}},
    \end{align}
where the new parameters $\varrho$ and $\xi$ are defined as follows:
\begin{align}
    \varrho&=\sqrt{\beta_+\beta_-}\\
    \xi&=\frac{\gamma_2}{\gamma_1+\varrho}
\end{align}
The entanglement entropy of the system in the von Neumann description is given below:
\begin{align}
    S(\xi)&=-\sum_{n=1}^\infty p_n(\xi)\ln{p_n(\xi)}\nonumber\\
    &=-\ln{(1-\xi)}-\frac{\xi}{1-\xi}\ln{\xi} \, ,
\end{align}
where $\xi$ is given by 
\begin{equation}
    \xi(R) =  \frac{(1-R)^2}{1+R^2+6R+4(1+R)\sqrt{R}}.
\end{equation}
and 
\begin{equation}
R \equiv  \frac{\beta_-}{\beta_+} = 
\sqrt{1-2 \left(\frac{\omega_1}{\omega_0}\right)^2} \, .
\end{equation}
$R$ takes values between $[0, 1]$.  In the decoupled limit ($\omega_1 \to 0$), $R \to 1$. In the limit of $\omega_1 \to \omega_0/\sqrt{2}$, $R \to  0$.

\begin{figure}[!hbt]
\centering
\includegraphics[scale=0.6]{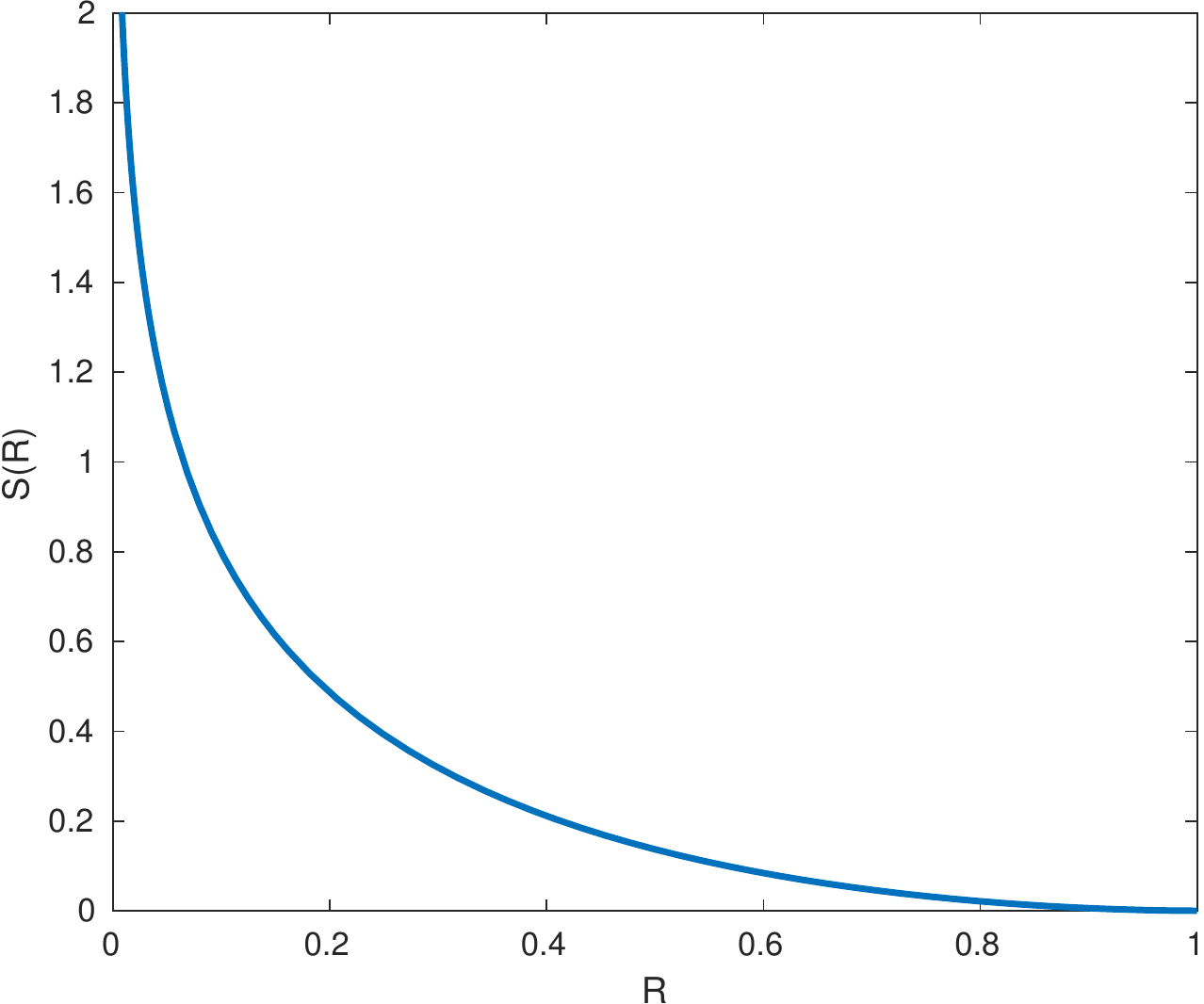}
\caption{Behaviour of entanglement entropy $S(R)$ for a coupled harmonic oscillator system, with respect to $R$.}
\label{fig1}
\end{figure}

We would like to point out the difference between the results in this section and that obtained by Srednicki~\cite{PhysRevLett.71.666}. In the case of a positive coupling constant,  the entanglement entropy diverges in the limit $\omega_-\to\infty$, which occurs when $\omega_1\to\infty$ (infinite coupling) and therefore corresponds to the UV divergence of the entropy. However, in our case (negative coupling constant), in the limit of $\omega_0 \to \sqrt{2}\omega_1$, $S \to \infty$. The divergence of the entanglement entropy occurs in the limit of $\omega_-\to 0$ i. e., in the zero frequency limit of \{$x_-$\} oscillator. Thus, the divergence is of IR origin with respect to the normal co-ordinates, even though this may not coincide with the zero energy limit in the original co-ordinates. This inversion in the nature of divergence of entropy is exactly captured by the UV/IR limits of systems in the normal coordinates, and this will further be clear from the scalar field model studied in the Section \hyperref[8]{VIII}.

To gain better insight into the above results, we need to analyze the behaviour of entropy as the normal mode oscillator approaches zero frequency, and for that purpose, we will develop the free particle approximation of harmonic oscillator in the next section.

\section{Free Particle Approximation Of Harmonic Oscillator}\label{3}

The analysis in the previous section demands the study of harmonic oscillator in the free particle limit. In this section, we show that we can achieve free particle approximation by taking $n\to\infty$ and $\omega\to0$ limits such that Energy is a constant. 

To go about doing that we begin with the WKB approximation of a harmonic oscillator:
\begin{equation}
    \Psi_n(x)=\frac{c}{\sqrt{p(x)}}\exp{\frac{i}{\hbar}\int p(x)d(x)}
\end{equation}
where $p(x)=\sqrt{2m(E_n-V(x))}$, and $V(x)=\frac{1}{2}m\omega^2x^2$. 
For a harmonic oscillator, the energy eigenvalues are 
$E_n=\hbar\omega(n+\frac{1}{2})$. As mentioned earlier, we take the 
following limits 
$$\lim_{\omega\to0}\lim_{n\to\infty}E_n\to E \, ,$$
i.e., energy eigenvalue tends to a constant $E$. In the rest of this section, we will keep $E$ to be a constant value; we will fix the value of $E$ in Section \hyperref[4]{\ref{4}} and Section \hyperref[5]{\ref{5}}. 

The turning points are given by $u=\pm\sqrt{2E_n/m\omega^2}$, and  we can see that $\lim_{\omega\to0}\lim_{n\to\infty}u\to\pm\infty$. The potential well in this limit is therefore stretched across the entire space, and the oscillator wave-function starts behaving like a plane wave. This can be seen from the fact that:

\begin{align}
\label{eq:HOtoFreeParticleLimit}
    &\lim_{\omega\to0}\lim_{n\to\infty}p(x)\to\sqrt{2mE} \\
  &\lim_{\omega\to0}\lim_{n\to\infty}\int p(x)dx\to\sqrt{2mE}x\nonumber
\end{align}
The resultant wave-function becomes:
\begin{equation}\label{plw}
   \Psi(x)=\frac{1}{\sqrt{\hbar k}}\big(c_1\exp{ikx}+c_2\exp{-ikx}\big),
\end{equation}
where $\hbar k=\sqrt{2mE}$. The wave-function (\ref{plw}) is that of a plane wave, and we may assume the boundary conditions as we seem fit. For convenience, we will only consider the incoming wave ($\Psi(x=\infty)=0$). Now, the normalization constant $c_1$ can be calculated from the condition $\int \Psi^*(x)\Psi(x)dx=1$, where the limits of integration are the classical turning points {$\pm\abs{u}$} prior to the free-particle approximation. This constant is found to be $c_1=\sqrt{m\omega/\pi}$. So we may write our final wave-function as follows:
\begin{equation}
    \Psi(x)=\frac{\exp{ikx}}{\sqrt{\Omega}},
\end{equation}
where $\Omega=\pi\hbar k/m\omega$ spans the volume of the space we are dealing with. It can be seen that the mapping is most physical when $\omega\to0$, and in this limit, $\Omega\to\infty$, which is expected. However, we will assume a negligible, positive value for $\omega$ in the above wave-function, and impose the physical limit at the end of our calculations (which can also be done by taking the $\Omega\to\infty$ limit).

\section{Zero Mode Analysis of Coupled Harmonic Oscillator}
\label{4}

In this section, we re-analyze the behaviour of entanglement entropy of a coupled harmonic oscillator, in the free particle limit of the $\{x_-\}$ oscillator. The calculations in Section \hyperref[2]{\ref{2}} were based on the ground state wave-function of the total system. The most  general wave-function in normal mode coordinate is given by: 
\begin{equation}
\Psi_{n_+,n_-}(x_+,x_-)=\psi_{n_+}(x_+)\psi_{n_-}(x_-) \, .
\end{equation}
Let us consider the case  where $n_+=0$, and take the limits 
$n_-\to\infty,   \omega_-\to0$. Using the results of the previous section, the above wave-function becomes:
\begin{equation}
    \Psi(x_+,x_-) \sim \frac{\beta_+^{1/4}}{\sqrt{\pi\Omega}}\exp{-\frac{\beta_+x_+^2}{2}+ik_-x_-},
\end{equation}
where 
\begin{equation}
k_-= \frac{\sqrt{2mE_-}}{\hbar}\, ; \quad \Omega=\frac{\pi\hbar k_-}{m\omega_-} \, .
\end{equation}
To obtain the entanglement entropy and understand the divergent behaviour, we rewrite the above wave-function in the physical coordinates, i. e., 
\begin{equation}
    \Psi(x_1,x_2)=\frac{\beta_+^{1/4}}{\sqrt{\pi\Omega}}\exp{-\frac{\beta_+(x_1+x_2)^2}{4}+i\frac{k_-(x_1-x_2)}{\sqrt{2}}}
\end{equation}
Following the procedure as in Section \hyperref[2]{II}, the reduced density matrix for the $x_1$ sub-system is:
{\small
\begin{align}
    \rho_1(x_1,x_1')
    &%=\int\limits_{-\infty}^{\infty}dx_2\Psi^*(x_1',x_2)\Psi(x_1,x_2)\nonumber\\
 =\frac{\sqrt{2}}{\Omega}\exp{-\frac{\beta_+(x_1-x_1')^2}{8}+i\frac{k_-(x_1-x_1')}{\sqrt{2}}}
\end{align}
}
From the above expression, it is clear that the reduced density matrix is homogeneous in space, i.e., $\rho_1(x_1,x_1')=\rho_1(x_1-x_1')$. The eigenvalues of $\rho_1(x_1,x_1')$ can be found in the Fourier domain:
\begin{equation}\label{fourier}
\!\!\!\int dx_1'\rho_1(x_1-x_1')\exp{-ikx_1'}=\tilde{\rho}_1(k)\exp{-ikx_1}
\end{equation}
The above integral equation tells us that $\tilde{\rho}_1(k)$, which is the Fourier transform of $\rho_1(x_1-x_1')$, spans the eigen-space of the latter. The eigenvalues are:
\begin{equation}
    \tilde{\rho}_1(k)=\frac{4}{\Omega}\sqrt{\frac{\pi}{\beta_+}}\exp{-\frac{(\sqrt{2}k-k_-)^2}{\beta_-}}
\end{equation}
Since $k$ is continuous, the entanglement entropy can be written in the integral form as:
\begin{equation}
    S=-\int\limits_{-\infty}^{\infty}\frac{dk}{(2\pi/\Omega)}\tilde{\rho}_1(k)\ln{\tilde{\rho}_1(k)},
\end{equation}
where $(2\pi/\Omega)$ is the volume of the $k$-space. We then get, 
\begin{equation}
    S=-\sqrt{2}\ln{\bigg(\frac{4\omega_-}{ek_-}\sqrt{\frac{m}{\pi\hbar\omega_+}}\bigg)},
\end{equation}
where $e$ is the Euler number.  The rest of the analysis is to have an understanding of the behavior of  the  entropy in the free-particle limit of $x_-$ oscillator i. e., $\omega_-\to0$. 

As mentioned in the previous section, we are free to choose the energy $E$, and consequently the wave number $k$ of the plane wave. There are two cases for the free particle limit $\omega_-\to0$:
\begin{itemize}
    \item \textbf{$\omega_0\to0$ and $\omega_1\to0$}: In this case, both $(x_+, x_-)$ oscillators are taken in the free particle limit. The limit is only valid if $\omega_1\to0$ is taken first and we later apply the condition $\omega_0\to0$. On matching the behaviour of entropy with that observed in Section \hyperref[2]{\ref{2}}, we can fix the plane wave energy to be
\begin{equation} 
\label{eq:IRlimit}   
    E_-= \frac{8\hbar\omega_0}{\pi e^2} \, . 
\end{equation}    
This will ensure that the log term vanishes in the first limit and that the behaviour of entropy is preserved in this approximation for this particular value of plane wave energy.
    \item \textbf{$\omega_0=\sqrt{2}\omega_1$}: In this case, only the oscillator $ x_-$ oscillator is taken in the free particle limit. The log term diverges negatively and, hence,  $S\to\infty$. Using the same plane wave energy as used in the previous case, we see that $E_-=(8/\pi e^2)\hbar\omega_0=(8\sqrt{2}/\pi e^2)\hbar\omega_1$.
\end{itemize}
\begin{figure}[htp]
\centering
\includegraphics[scale=0.6]{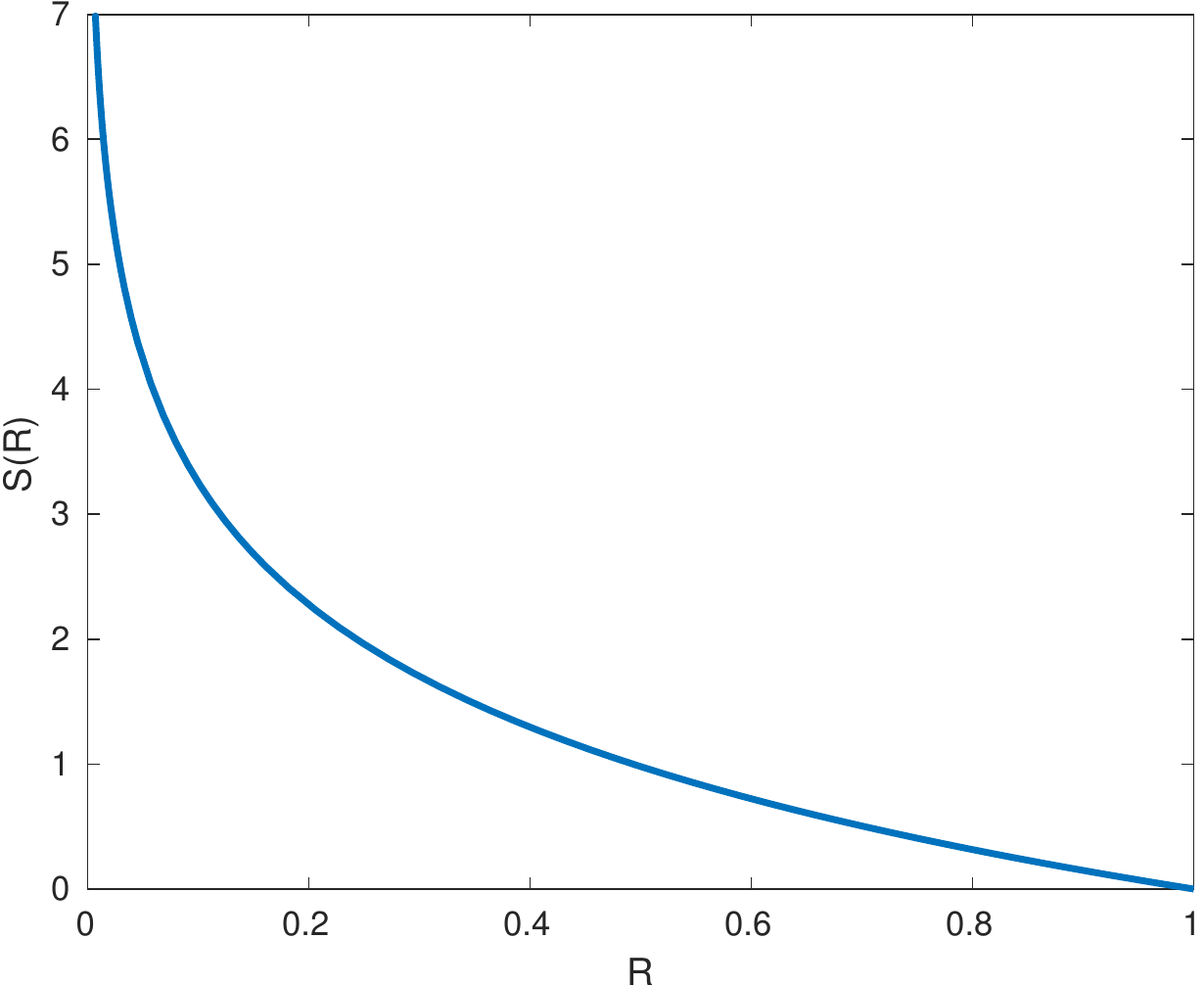}
\caption{Plot of entanglement entropy with respect to $R$, where $R=\omega_-/\omega_+$, and $S(R)=-\sqrt{2}\ln{R}$.}
\label{fig2}
\end{figure}

The results obtained in this section are consistent with those obtained in Section \hyperref[2]{\ref{2}}, and  confirm that the divergence of entropy occurs in the zero-mode limit of the system in normal co-ordinates. The procedure developed in this section can be used for any oscillator system in the IR regime of its normal coordinates. We will explore this in the rest of this work.

%Moreover, in oscillator systems that involved positive coupling, these divergences were observed to occur due to the accumulation of a large number of near-zero-modes\cite{PhysRevD.90.044058}, where each such mode could be associated with an undetermined non-zero energy\cite{PhysRevD.39.3642}. The IR divergences in this section, however, occur due to a single zero-mode associated with a finite, positive energy (Case 2), and this is exclusive to negative coupling.

\section{Hydrogen atom as a bipartite system}
\label{5}

We will now look at the case of a Hydrogen atom --- an entangled system of  electron and proton sub-systems. Such an analysis has been performed earlier in Ref.~\cite{doi:10.1119/1.18977}; however, here we will build on a new perspective that will treat Hydrogen atom as a limiting case of a broader system.

Consider the Hamiltonian of the Hydrogen atom:
 \begin{equation}
 \label{eq:HydroHamitonian1}
 H_{\rm Hyd}=\frac{p_e^2}{2m_e}+\frac{p_p^2}{2m_p}+\frac{e^2}{\abs{\vec{r_e}-\vec{r_p}}}.
 \end{equation}
On performing the transformations 
\begin{equation}
\vec{R}=(m_e\vec{r_e}+m_p\vec{r_p})/M;~~\vec{r}=\vec{r_e}-\vec{r_p} \, ,
\end{equation}
 the Hamiltonian becomes: 
\begin{equation}
\label{eq:HydroHamitonian}
H_{\rm Hyd} =H_{\rm cm}+H_{\rm int};~~H_{\rm cm}=\frac{p_R^2}{2M};~~
H_{\rm int}=\frac{p_r^2}{2m} -\frac{e^2}{r}
\end{equation} 
where $M=m_e+m_p$ is the total mass and $m=m_em_p/(m_e+m_p)$ is the reduced mass. Thus, the Hamiltonian gets decoupled in 
$(\vec{R}, \vec{r})$ coordinates. The resulting wave-function is a product of the eigenstates of $H_{\rm cm}$ and $H_{\rm int}$:
 \begin{equation}
     \Psi(\vec{R},\vec{r})=\phi_{int}(\vec{r})\frac{\exp{i\vec{P}.\vec{R}/\hbar}}{\sqrt{\Omega}}
\label{eq:HyrdoPsi} 
 \end{equation}
In the previous section, we showed that the free particle can be treated as a limit of Harmonic oscillator. In other words, the Hydrogen atom Hamiltonian can be treated as limit of the following Hamiltonian:
 \begin{equation}
 \label{eq:Gen-HydroHamiltonian}
 H=\frac{p_R^2}{2M}+\frac{p_r^2}{2m}+\frac{1}{2}M\omega^2R^2-\frac{e^2}{r}
 \end{equation}
The above Hamiltonian has two different coupling parameters ---  oscillator frequency  $(\omega)$, and electronic charge $(e)$. 
Naturally, this system is entangled in ($\vec{r_e},\vec{r_p}$) 
co-ordinates and $(\omega, e)$ contain information about the extent to which the $\vec{r_e}$ and $\vec{r_p}$ sub-systems are entangled.
Thus, the Hydrogen atom system is the free-particle approximation of a broader system which consists of a 3D harmonic oscillator and a 3D Coulomb sub-system.  

The advantages of using Hamiltonian (\ref{eq:Gen-HydroHamiltonian}) instead of the Hydrogen atom Hamiltonian (\ref{eq:HydroHamitonian}) is two-fold: first, as discussed in Section \hyperref[3]{\ref{3}},  the arbitrariness of $\Omega$ is removed. In the case of Hamiltonian (\ref{eq:HydroHamitonian}), the centre-of-mass sub-system is non-normalizable. Second,  we will be able to quantify the divergence of the entanglement entropy precisely. 
 
For simplicity, we assume that the energy of the free particle ($P^2/2M$) is equally distributed among these oscillators and the corresponding momenta are given by $P_x=P_y=P_z=P/\sqrt{3}$.  Using the harmonic oscillator to free particle limit (\ref{eq:HOtoFreeParticleLimit}), the centre-of-mass wave-function can be written as:
\begin{multline}
    \varphi(X,Y,Z)=\frac{\exp{iPX/\sqrt{3}\hbar}}{\sqrt{\Omega_X}}\frac{\exp{iPY/\sqrt{3}\hbar}}{\sqrt{\Omega_Y}}\\\times\frac{\exp{iPZ/\sqrt{3}\hbar}}{\sqrt{\Omega_Z}},
\end{multline} 
where $\vec{R}=(X,Y,Z)$. These plane waves are the limiting cases of three identical and mutually perpendicular 1D oscillators that occupy the respective axes of three-dimensional space. Using (\ref{eq:IRlimit}), the normalization volume $\Omega$ in this limit is:
\begin{equation}
    \Omega=\Omega_X\Omega_Y\Omega_Z=\bigg(\frac{\pi P}{\sqrt{3}M\omega}\bigg)^3 \, .
\end{equation} 
It is important to note that $\omega\to0$ corresponds to 
$\Omega\to\infty$. With this background, we now proceed to evaluate  entanglement entropy for the Hydrogen atom. In the ground state, the internal wave-function $\phi_{int}$ is given by:
 \begin{equation}
     \phi_{int}(r)=\frac{1}{\sqrt{\pi}}\bigg(\frac{1}{a_0}\bigg)^{3/2}\exp{-\frac{r}{a_0}},
 \end{equation}
 where the Bohr radius is given by $a_0=\hbar^2/me^2$. In terms of the  electron and proton coordinates, the above wave-function is:
\begin{equation}
    \Psi(\vec{r_e},\vec{r_p})=\frac{1}{\sqrt{\Omega}}\phi_{int}(\vec{r_e}-\vec{r_p})\exp{\frac{i}{\hbar}\vec{P}.\frac{m_e\vec{r_e}+m_p\vec{r_p}}{M}}
\end{equation}
This is an entangled state. In order to obtain the reduced density matrix for the electron system, we trace over the proton basis set of co-ordinate eigenfunctions:
\begin{equation}
\begin{split}
\rho(\vec{r_e},\vec{r_e'}) 
&=\int d^3\vec{r_p}\Psi^*(\vec{r_e'},\vec{r_p})\Psi(\vec{r_e},\vec{r_p})\\
&=\frac{1}{\Omega}\exp{i\frac{m_e}{\hbar M}\vec{P}.(\vec{r_e}-\vec{r_e'})}\\&\qquad\qquad\times\int d^3\vec{r_p}\phi^*_{int}(\vec{r_e'}-\vec{r_p})\phi_{int}(\vec{r_e}-\vec{r_p})\\
&=\exp{i\frac{m_e}{\hbar M}\vec{P}.(\vec{r_e}-\vec{r_e'})}\rho_{int}(\vec{r_e},\vec{r_e'}),
\end{split}
\end{equation}
where $\rho_{int}$ is the density matrix of the electron sub-system when the atom is at rest ($\vec{P}=0$). Note that the reduced density matrix is homogeneous in space. To prove this, it suffices to show that $\rho_{int}(\vec{r_e},\vec{r_e'})$ is a function of $\vec{r_e}-\vec{r_e'}$. To this end, we define a new vector $\vec{y}=\vec{r_e'}-\vec{r_p}$ and rewrite $\rho_{int}$ as follows:
\begin{equation}
    \rho_{int}(\vec{r_e},\vec{r_e'})=\int{\frac{d^3\vec{y}}{\Omega\pi a_0^3}\exp{-\frac{\abs{\vec{r_e}-\vec{r_e'}+\vec{y}}}{a_0}}}\exp{-\frac{\abs{\vec{y}}}{a_0}}
\end{equation}
Like in Eq. (\ref{fourier}), using the Fourier Transform  of $\rho_{int}(\vec{r_e}-\vec{r_e'})$, we get:
\begin{equation}
    \int d^3\vec{r_e'}\rho_{int}(\vec{r_e}-\vec{r_e'})\exp{i\vec{k}.\vec{r_e'}}=\tilde{\rho}_{int}(\vec{k})\exp{i\vec{k}.\vec{r_e}}
\end{equation}
From the above integral equation, it can be seen that the eigenvalues of the reduced density matrix are given by $\tilde{\rho}_{int}(\vec{k})$. For the $1s$ electron in an atom moving with momentum $\vec{P}$, the eigenvalues of the reduced density matrix $\rho(\vec{r_e}-\vec{r_e'})$ are found to be: 
\begin{equation}\label{eig1}
    \tilde{\rho}(\vec{k})=\tilde{\rho}_{int}(\vec{k}-\frac{m_e\vec{P}}{M\hbar })=\frac{1}{\Omega}\frac{64\pi a_0^3}{\big(1+a_0^2\abs{\vec{k}-\vec{k_e}}^2\big)^4},
\end{equation}
where $\vec{k_e}=m_e\vec{P}/M\hbar$.  The entanglement entropy of the system is given by 
\begin{equation}
    S=-\int  d^3\vec{k} \frac{\Omega}{(2\pi)^3}\tilde{\rho}(\vec{k})\ln{\tilde{\rho}(\vec{k})}
\end{equation}
Note that the above expression is dimensionless, since $(2\pi)^3/\Omega$ spans the volume of the $\vec{k}$-space. Rewriting the integral as 
$$S=\int\limits_0^\infty d\kappa \, g(\kappa) \, , $$
where $\kappa=k/k_e$ is dimensionless, $\eta=a_0 \, k_e$, $\zeta=64\pi a_0^3/\Omega$, $c_0 = 16\eta^3\kappa^2/{\pi}$ and 

%\begin{widetext}
{\small
\begin{multline}\label{g}
\!\!\!\! g(\kappa)=- c_0 \Bigg\{\Bigg[\frac{1}{\big(1+\eta^2(1-\kappa)^2\big)^4}- \frac{1}{\big(1+\eta^2(1+\kappa)^2\big)^4}\Bigg] \ln{\zeta} \\ 
+ \frac{1}{\big(1+\eta^2(1-\kappa)^2\big)^4}\ln{\bigg(\frac{1}{\big(1+\eta^2(1-\kappa)^2\big)^4}\bigg)}\\-\frac{1}{\big(1+\eta^2(1+\kappa)^2\big)^4}\ln{\bigg(\frac{1}{\big(1+\eta^2(1+\kappa)^2\big)^4}\bigg)}\Bigg\}
\end{multline}
}
%\end{widetext}
%
\begin{figure}[ht!]
    \centering
    \includegraphics[scale=0.6]{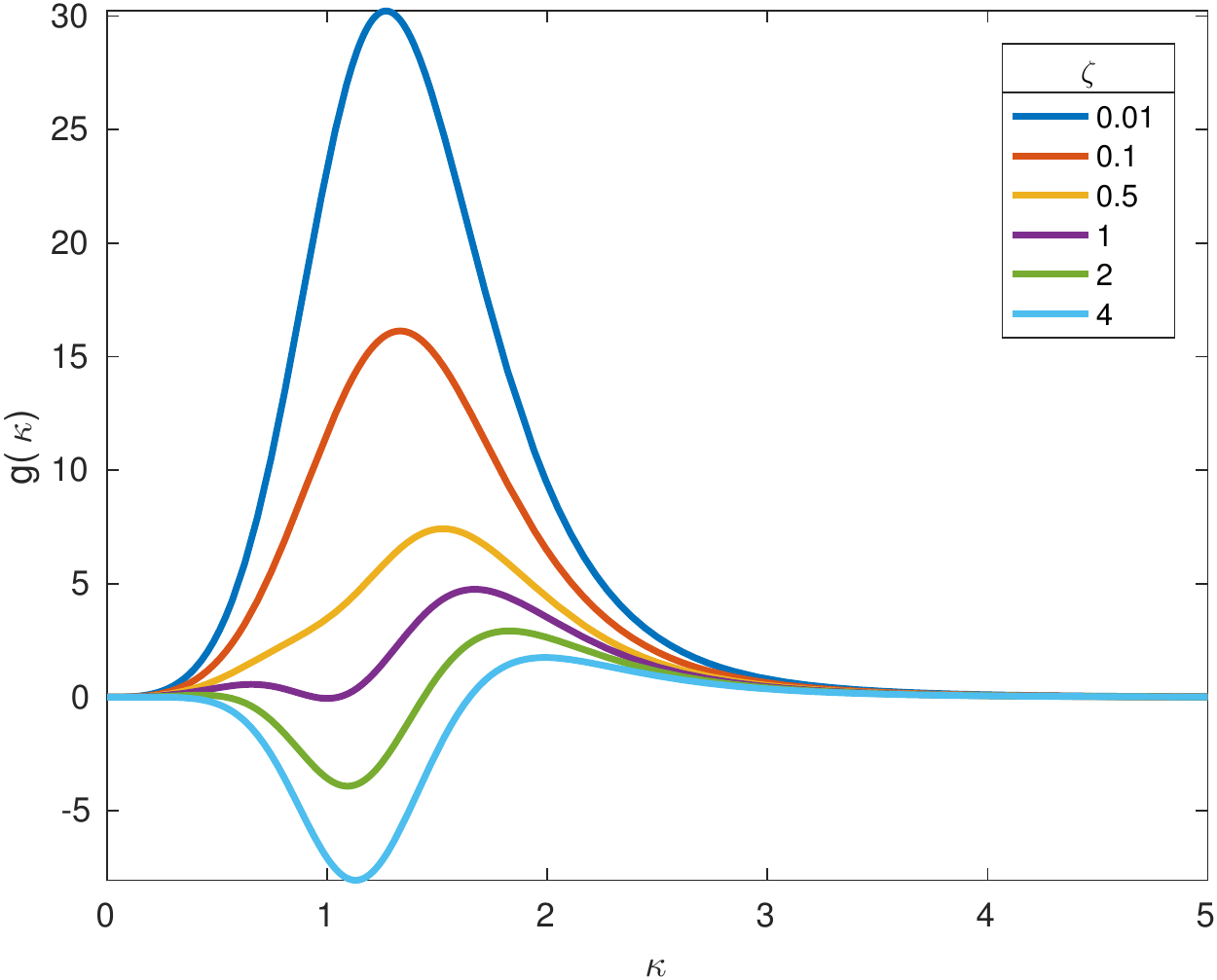}
    \caption{Plot of $g(\kappa)$ with respect to $\kappa$ for different values of $\zeta$ (where $\zeta\propto\omega^3$), given the initial condition $\eta=1$. Note that the entropy, given by the area of $g(\kappa)$, is positive for a sufficiently small $\zeta$, and monotonically increases for smaller values of $\zeta$. }
    \label{fig3}
\end{figure}

Although an analytical expression for the entropy may be complicated, we can deduce the behaviour of entropy by analyzing the area of $g(\kappa)$. From the above figure, we observe the following: First, $\lim_{\kappa\to0}g(\kappa)=\lim_{\kappa\to\infty}g(\kappa)=0$.  Second, for sufficiently small $\zeta$, the area of $g(\kappa)$ is positive and finite. Third, for $\kappa>0$, $\eta^2(1+\kappa)^2>\eta^2(1-\kappa)^2$ in (\ref{g}). Thus, when the Hamiltonian (\ref{eq:Gen-HydroHamiltonian}) reduces to Hydrogen atom or in the zero-mode limit ($\omega\to0$), $\zeta\to0$. Thus, $g(\kappa)$ diverges Logarithmically and consequently, the entropy (which is the area of $g(\kappa)$) diverges positively.  The entanglement entropy of the Hydrogen atom exhibits zero-mode divergence.

\section{Coulomb Problem-Oscillator Mapping in Hydrogen Atom}\label{6}

In the previous section, we explicitly showed that the entanglement entropy of Hydrogen atom exhibits zero-mode divergence. In this section, we show that this behaviour remains the same under certain transformations. It is well-known that a three-dimensional Hydrogen atom can be mapped to a constrained four-dimensional harmonic oscillator~\cite{doi:10.1119/1.17065}. We use this transformation and obtain entanglement entropy in the new coordinate system, and subsequently observe the zero-mode divergence of entanglement entropy.  

Our starting point is the Hamiltonian (\ref{eq:HydroHamitonian1}) in  $(\vec{r},\vec{R})$ coordinates. Using the wave-function (\ref{eq:HyrdoPsi}), 
the Schrodinger equation splits into two differential equations:
{\small
\begin{eqnarray}
\label{eq:HydroSchr}
& & \left[-\frac{\hbar^2}{2 \, m}\nabla_r^2-\frac{e^2}{r} \right] \phi_{int} 
= - B \phi_{int};
%\nonumber \\
%
%& &
-\frac{\hbar^2}{2M}\nabla_R^2\varphi=\frac{P^2}{2M} \varphi
\end{eqnarray}
}
where $\vec{P}$ is the momentum of the centre of mass sub-system and $B=-E+P^2/2M$ is the binding energy. 

To map the hydrogen atom system to set of harmonic oscillators, we rewrite the Hamiltonian in terms of complex variables~\cite{doi:10.1119/1.17065}: 
\begin{eqnarray}
\xi_1 &=& \sqrt{r}\cos{(\theta/2)}\exp{i\nu_1/2} \nonumber \\
\xi_2 &=& \sqrt{r}\sin{(\theta/2)}\exp{i\nu_2/2}
\end{eqnarray}
where $\nu_1=\sigma+\phi$ and $\nu_2=\sigma-\phi$. Note that 
$\sigma$ is as an extra variable in the mapping.

The Hamiltonian (\ref{eq:HydroSchr}) can then be rearranged to obtain a system of two complex harmonic oscillators\cite{0295-5075-98-3-30008}:
\begin{equation}
    \bigg\{-\frac{\hbar^2}{2m}\bigg(\frac{\partial^2}{\partial\xi_1^*\partial\xi_1}+\frac{\partial^2}{\partial \xi_2^*\partial\xi_2}\bigg)+B(\xi_1^*\xi_1+\xi_2^*\xi_2)\bigg\}\phi_{int}=e^2\phi_{int}
\end{equation}
Demanding that the wave-function should be independent of $\sigma$ ensures that the degrees of freedom are conserved. The constraint equation $\partial\phi_{int}/\partial\sigma=0$ leads to:
\begin{equation}\label{constraint}
    \bigg(\xi_1^*\frac{\partial\phi_{int}}{\partial\xi_1^*}-\xi_1\frac{\partial\phi_{int}}{\partial\xi_1}\bigg)=-\bigg(\xi_2^*\frac{\partial\phi_{int}}{\partial\xi_2^*}-\xi_2\frac{\partial\phi_{int}}{\partial\xi_2}\bigg)
\end{equation}
We can now define the creation and annihilation operators:
\begin{align}
a_+&=-\frac{i}{\sqrt{2}}\bigg(\frac{\hbar}{\sqrt{2Bm}}\bigg)^{1/2}\bigg(\frac{\partial}{\partial\xi_1}+\frac{\sqrt{2mB}}{\hbar}\xi_1^*\bigg)\nonumber\\
a_-&=-\frac{i}{\sqrt{2}}\bigg(\frac{\hbar}{\sqrt{2Bm}}\bigg)^{1/2}\bigg(\frac{\partial}{\partial\xi_1^*}+\frac{\sqrt{2mB}}{\hbar}\xi_1\bigg)\nonumber\\
a_+^\dagger&=-\frac{i}{\sqrt{2}}\bigg(\frac{\hbar}{\sqrt{2Bm}}\bigg)^{1/2}\bigg(-\frac{\partial}{\partial\xi_1^*}+\frac{\sqrt{2mB}}{\hbar}\xi_1\bigg)\nonumber\\
a_-^\dagger&=-\frac{i}{\sqrt{2}}\bigg(\frac{\hbar}{\sqrt{2Bm}}\bigg)^{1/2}\bigg(-\frac{\partial}{\partial\xi_1}+\frac{\sqrt{2mB}}{\hbar}\xi_1^*\bigg) \, .
\end{align}
Similarly, the creation and annihilation operators $b_\pm$ are obtained from the above set of equations by replacing $\xi_1$ everywhere with $\xi_2$. The Schr\"{o}dinger equation in terms of the ladder operators then takes the form of a 4D harmonic oscillator:
\begin{equation}\label{harm}
    \hbar\sqrt{B/2m}(a_+^\dagger a_++a_-^\dagger a_-+b_+^\dagger b_++b_-^\dagger b_-+2)\phi_{int}=e^2\phi_{int}
\end{equation}
Referring to the Schwinger representation for a harmonic oscillator, we can define the angular momenta of the sub-systems (labelled here as $a$ and $b$) as $J_z^{(a)}=a_+^\dagger a_+-a_-^\dagger a_-$ and $J_z^{(b)}=b_+^\dagger b_+-b_-^\dagger b_-$. Taking into account the expression in (\ref{constraint}), we see that $(J_z^{(a)}+J_z^{(b)})\phi_{int}=0$. This constraint clearly arises from the fact that the wave-function $\phi_{int}$ is independent of $\sigma$. Consequently, (\ref{harm}) takes the form:
\begin{equation}\label{eig}
    \hbar\tilde{\omega}(a_+^\dagger a_++b_+^\dagger b_++1)\phi_{int}=e^2\phi_{int},
\end{equation}
where $\tilde{\omega}=\sqrt{2B/m}$. On rearranging the energy eigenvalue equation, we get $B=me^4/2n^2\hbar^2$ (the well known Rydberg formula) where $n=n_++m_++1$. If we look at the ground state ($n=1$) of 3D Coulomb sub-system, we find that $n_+=m_+=0$, which implies that it corresponds to the ground state of the 4D harmonic oscillator. 

The ground state wave function in the transformed coordinates is given by:
\begin{equation}\label{oscwave}
\!\!\!\! \phi_{int}(\abs{\xi_1},\abs{\xi_2},\nu_1,\nu_2)=C\exp{-\frac{m\tilde{\omega}}{\hbar}(\abs{\xi_1}^2+\abs{\xi_2}^2)}
\end{equation}
 Keeping in mind that the degrees of freedom are conserved in the transformation, using the Jacobian, we may write the normalization condition as
\begin{multline}\label{norm1}
   1=8 \int\limits_0^\infty d\abs{\xi_1}\int\limits_0^\infty d\abs{\xi_2}\int\limits_0^{2\pi}d\phi\,\abs{\xi_1}\abs{\xi_2}(\abs{\xi_1}^2+\abs{\xi_2}^2)\\\times\abs{\phi_{int}(\abs{\xi_1},\abs{\xi_2},\phi)}^2
\end{multline}
where $C=\beta^{3/2}/\sqrt{8\pi}$. Despite the oscillator system being uncoupled, we see that there is a non-zero entanglement in the oscillator co-ordinates with respect the Hydrogen atom system. See \hyperref[app:a]{Appendix \ref{app:a}} for details. Returning to the original $\vec{r_e},\vec{r_p}$ coordinates, $\Psi$ is given by:
\begin{multline}
    \Psi(\vec{r_e},\vec{r_p})=\frac{\beta^{3/2}}{\sqrt{8\pi\Omega}}\exp{-\frac{\beta}{2}\abs{\vec{r_e}-\vec{r_p}}}\\\times\exp{i\frac{\vec{P}}{\hbar}.\frac{m_e\vec{r_e}+m_p\vec{r_p}}{M}}
\end{multline}
The reduced density matrix for the electronic sub-system is of the form:
\begin{equation}
\rho(\vec{r_e},\vec{r_e'})=\int d^3\vec{r_p}\Psi^*(\vec{r_e'},\vec{r_p})\Psi(\vec{r_e},\vec{r_p})
\end{equation}
On proceeding to find the eigenvalues of reduced density matrix as was done in Section \hyperref[5]{V}, we get:
\begin{equation}
    \tilde{\rho}(\vec{k})=\frac{1}{\Omega}\frac{64\pi a_0^3}{\big(1+a_0^2\abs{\vec{k}-\vec{k_e}}^2\big)^4}
\end{equation}
We find that the eigenvalues obtained have the same form as in the unmapped case (\ref{eig1}). The behaviour of entropy is therefore exactly preserved, and a similar result is also obtained in the Coulomb problem-isotonic oscillator mapping (\hyperref[app:b]{Appendix \ref{app:b}}). The entropy diverges when $\Omega\to\infty$ for both cases, and this implies that entanglement entropy under such transformations of the Hamiltonian retains its property of zero-mode divergence.

\section{Zero-mode divergence of environment-induced entanglement}\label{7}
Having established zero-mode divergence of entanglement entropy in the Hydrogen atom and coupled Harmonic oscillators in the previous sections, we now look at a slightly different scenario \cite{2002-Braun-PRL,*2013-Ghesquiere.etal-PLA}. Consider a case where two harmonic oscillators $\{x_1,x_2\}$, that do not interact with each other, however, interact with the environment which is represented by the oscillator $\{y\}$. Such a system generates correlations between oscillators $x_1$ and $x_2$,
due to the oscillator $y$, the nature of which is quite unlike our usual definitions of quantum entanglement. The Hamiltonian for such a system is given below:
 \begin{equation}
     H=\sum_{i=1}^2\bigg[\frac{p_i^2}{2m}+\frac{1}{2}m\omega^2x_i^2\bigg]+\frac{p_y^2}{2M}+\frac{1}{2}M\Omega^2y^2+\alpha x_1y+\beta x_2y
 \end{equation}
We will now scale the Hamiltonian as $\tilde{H}=H/\omega$ and make the transformations $x_{i}=\tilde{x}_{i}/\sqrt{m\omega}$, $y=\tilde{x_3}/\sqrt{M\omega}$, $\alpha=\tilde{\alpha}\sqrt{Mm}\omega^2$, $\beta=\tilde{\beta}\sqrt{Mm}\omega^2$ and $\Omega^2=k\omega^2$:
\begin{equation}
    \tilde{H}=\sum_{i=1}^3\frac{\tilde{p}_i^2}{2}+\sum_{i,j=1}^3\frac{1}{2}\tilde{x}_iK_{ij}\tilde{x}_j,
\end{equation}
where the matrix $\{K_{ij}\}$ has the form:
\begin{equation}
    K=\begin{bmatrix}1&0&\tilde{\alpha}\\0&1&\tilde{\beta}\\\tilde{\alpha}&\tilde{\beta}&k\end{bmatrix}
\end{equation}
The eigenvalues of $K$ are given by:
\begin{align}
\kappa_1&=1\nonumber\\
\kappa_2&=\frac{1}{2}\left(1+k-\sqrt{(k-1)^2+4(\tilde{\alpha}^2+\tilde{\beta}^2)}\right)\nonumber\\
\kappa_3&=\frac{1}{2}\left(1+k+\sqrt{(k-1)^2+4(\tilde{\alpha}^2+\tilde{\beta}^2)}\right)
\end{align}
From the above expressions, it is clear that the eigenvalues are always real. Since the eigenvalues are squared normal modes, we can have the following scenarios depending on their sign:
\begin{itemize}
    \item \textbf{Normal Mode Oscillator:}  The normal modes given by the eigenvalues are all positive. This corresponds to the condition $\tilde{\alpha}^2+\tilde{\beta}^2<k$. The coupling constants are therefore bounded.
    \item \textbf{Free Particle Case:} One of the eigenvalues ($\kappa_2$), and the corresponding normal mode becomes zero. The sub-system $\{y_2\}$ is now a free particle and not an oscillator. The condition for this case is $\tilde{\alpha}^2+\tilde{\beta}^2=k$.
    \item \textbf{Inverted Oscillator:} At least one eigenvalue becomes negative, and corresponds to the condition $\tilde{\alpha}^2+\tilde{\beta}^2>k$. In this case, the energy eigenvalues can be negative, as a result of which it is generally considered to be unphysical.
\end{itemize}
% The resultant Hamiltonian is therefore of the form:
% \begin{equation}
%     \tilde{H}=\sum_{i=1}^3\bigg[\frac{\tilde{p}_{\tilde{y}_i}^2}{2}+\frac{1}{2}\kappa_i\tilde{y}_i^2\bigg]
% \end{equation}
We are interested in the free-particle case $\kappa_2=0$ corresponding to $k=\tilde{\alpha}^2+\tilde{\beta}^2$. This is also equivalent to the expression:
\begin{equation}
    \alpha^2+\beta^2=M\Omega^2m\omega^2
\end{equation}
In this limit, the eigenvalues are:
\begin{align}
\kappa_1&=1;~~ \kappa_2 =0;~~
\kappa_3=1+\tilde{\alpha}^2+\tilde{\beta}^2
\end{align}
The normal mode co-ordinates in this limit are given by:
\begin{align}
    \tilde{z}_1&=\frac{-\tilde{\beta}\tilde{x}_1+\tilde{\alpha}\tilde{x}_2}{\sqrt{\tilde{\alpha}^2+\tilde{\beta}^2}}\nonumber\\
    \tilde{z}_2&=\frac{-\tilde{\alpha}\tilde{x}_1-\tilde{\beta}\tilde{x}_2+\tilde{x}_3}{\sqrt{1+\tilde{\alpha}^2+\tilde{\beta}^2}}\nonumber\\
    \tilde{z}_3&=\frac{\tilde{\alpha}\tilde{x}_1+\tilde{\beta}\tilde{x}_2+(\tilde{\alpha}^2+\tilde{\beta}^2)\tilde{x}_3}{\sqrt{(\tilde{\alpha}^2+\tilde{\beta}^2)(1+\tilde{\alpha}^2+\tilde{\beta}^2)}}
\end{align}
The Hamiltonian is now of the form:
\begin{equation}
    \tilde{H}=\sum_{i=1}^3\frac{\tilde{p}_{\tilde{z}_i}^2}{2}+\frac{1}{2}\tilde{z}_1^2+\frac{1}{2}(1+\tilde{\alpha}^2+\tilde{\beta}^2)\tilde{z}_3^2
\end{equation}
Extending the procedure for calculating entanglement entropy that was developed in Section \hyperref[2]{II} to $N$ coupled Harmonic oscillators~\cite{*PhysRevLett.71.666} (where $N=3$), we obtain $\xi=1$, and $S_1=\infty$. The divergence of entanglement entropy $S_1$ associated with the reduced density matrix of $\{x_1\}$ oscillator is therefore due to zero-modes in the normal co-ordinates, and the same result is obtained for the entropy $S_2$ associated with reduced density matrix of $\{x_2\}$ oscillator.

\section{Conclusions and Discussions}\label{8}

We have investigated, in detail, the origin of zero-mode divergences of entanglement entropy in simple quantum systems. We have shown explicitly that --- negatively-coupled harmonic oscillator (in Sections \hyperref[2]{II}, \hyperref[4]{IV}) and Hydrogen atom (in Sections \hyperref[5]{V}) and tri-partite oscillator system involving environment-induced entanglement (in Section \hyperref[7]{VII}) --- all exhibit zero-mode divergence of entanglement entropy. 

To quantify the divergence of entanglement entropy in the zero-mode limit, in Section \hyperref[3]{III}, we systematically developed the free-particle approximation of a harmonic oscillator. The approximation provided an unambiguous way to fix the energy for zero-modes and probe the divergence of entropy in the IR regime of the system, in the normal co-ordinates. In Section \hyperref[5]{V}, we were able to reverse this approximation to arrive at a broader system of which the Hydrogen atom is a particular case, and yet again the entropy was found to diverge in the zero-mode limit. Furthermore, in Section \hyperref[6]{VI}, we showed that the characteristic behaviour of entanglement entropy is preserved on mapping the Coulomb sub-system of a 3D Hydrogen atom to 4D harmonic oscillator suggesting that it possesses certain symmetries that the Hamiltonian does not. This also hints at a broader class of transformations that preserve the zero-mode divergence of entanglement entropy, which can be used to study more complex systems, and especially those that were initially thought to exhibit UV-divergence. We further observe zero-mode divergence in Section \hyperref[7]{VII} for the case of two uncoupled oscillators that interact with each other through a third oscillator.  

Given the above results, the question remains as to whether the zero-mode divergence is ubiquitous across physical systems. To go about understanding this, let us consider free, massive scalar field as studied in Ref. \cite{PhysRevD.90.044058}:
\begin{equation}\label{1d}
    H^{(\text{1D})}=\frac{1}{2}\sum_{n=0}^{N-1}\bigg(\frac{\pi_n^2}{a}+\frac{1}{a}(\phi_{n+1}-\phi_n)^2+am_f^2\phi_n^2\bigg)
\end{equation}
The continuum limit corresponds to $a\to0$ (where $a$ is the lattice spacing), and we have also assumed periodic boundary conditions ($\phi_0=\phi_N$). The normal modes in this case, keeping in mind $N\to\infty$, are given by:
\begin{equation}
    \omega_k^2=m_f^2+\frac{4}{a^2}\sin^2{\frac{\pi k}{N}},\:\:\:k \, = \, 0, 1,...,N-1
\end{equation}
Despite the coupling constant in (\ref{1d}) being negative, as was the case in Section \hyperref[2]{II}, the system exhibits UV divergence of entanglement entropy. This is because not even a single normal mode vanishes, unless we consider the massless case ($m_f=0$). The specific case for $N=3$ has been worked out in \hyperref[appc]{Appendix C}. Following Ref. \cite{PhysRevD.90.044058}, we now perform the following canonical transformation:
\begin{align}
    \pi_n&=\bar{\pi}_n(2+a^2m_f^2)^{1/4}\nonumber\\
    \phi_n&=\frac{\bar{\phi}_n}{(2+a^2m_f^2)^{1/4}}
\end{align}
The normal modes now become:
\begin{align}
    \bar{\omega}_i&=\sqrt{1-\mu+2\mu\sin^2{\frac{\pi i}{N}}}\\
    \mu&=\frac{2}{2+a^2m_f^2}
\end{align}
We obtain at least one zero-mode, corresponding to the case $i=0$ and $\mu\to1$. Consequently, the entanglement entropy was shown to diverge due to the accumulation of a large number of near-zero-modes, and at-least one zero-mode. This confirms that these divergences occur only when the system develops at least one zero-mode in the normal co-ordinates (in accordance with the results of this paper), and in all other cases the entanglement entropy diverges in the UV limit. It remains to be seen if these divergences occur in the IR limit of the system in its original co-ordinates, thus giving us a distinct criteria for UV-IR divergences of entanglement entropy. This will be further investigated in later work.

In the Introduction, we asked: How entanglement stores information about the possible experimental outcome without having information about the individual sub-systems? In this regard, our result suggests that the zero-modes of the entanglement entropy, corresponding to the long wave-length correlation, may provide insight into this. We hope to return to study this in the near future.

Our analysis may have implications in understanding the IR structure of gravity~\cite{2017-Strominger-Arx}\cite{PhysRevLett.119.180502}. The information paradox due to the Hawking radiation seems to be associated with the IR because of the infinite number of soft gravitons produced in the black-hole evaporation~\cite{2016-Hawking.etal-PRL}. The emitted photons are highly entangled and understanding the entanglement of these soft-photons may shine some light on the information paradox. We hope to address this elsewhere.

\begin{acknowledgements}
The authors thank Bindusar Sahoo for useful discussions. The work is supported by IRCC seed grant, IIT Bombay. SMC is currently supported by IRCC Seed Grant, IIT Bombay, and was previously supported by Inspire fellowship of DST, Government of India.  SS is partly supported by Homi Bhabha Fellowship Council.
\end{acknowledgements}
\appendix
\section{Entanglement in Oscillator coordinates}
\label{app:a}

Since the mapped wave-function in oscillator coordinates(\ref{oscwave}) is in an uncoupled form, we naturally expect the entropy to be zero. However, in this section, we show that there is an entanglement that develops in the mapped coordinates with respect to the original system. In order to avoid complications when we resort to binomial expansion later on in (\ref{bin}), we shift to a dimensionless Schr\"odinger equation by setting the units $e=m=\hbar=1$. Looking at (\ref{norm1}), we see that the partial trace integral is quite non-trivial in the oscillator coordinates. So we define an effective wave-function $\Phi(\abs{\xi_1},\abs{\xi_2})=2\sqrt{2\abs{\xi_1}\abs{\xi_2}(\abs{\xi_1}^2+\abs{\xi_2}^2)}\phi_{int}(\abs{\xi_1},\abs{\xi_2},\phi)$. Now the normalization condition in (\ref{norm1}) becomes:
\begin{equation}
    \int\limits_{-\infty}^{\infty} d\abs{\xi_1}\int\limits_{-\infty}^{\infty} d\abs{\xi_2}
    \int\limits_0^{2\pi}d\phi\,\abs{\Phi(\abs{\xi_1},\abs{\xi_2},\phi)}^2=1
\end{equation}
The normalized ground state wave-function in oscillator coordinates will be:
\begin{multline}
    \Phi(\abs{\xi_1},\abs{\xi_2},\phi)=\sqrt{\frac{\abs{\xi_1}\abs{\xi_2}(\abs{\xi_1}^2+\abs{\xi_2}^2)}{\pi}}\\\times\exp{-\frac{\abs{\xi_1}^2+\abs{\xi_2}^2}{2}}
\end{multline}
Now we may easily trace out any sub-system from the effective density matrix. The reduced density matrix of the $\{\abs{\xi_2}\}$ sub-system can be calculated as follows:
\begin{equation}\label{red}
    \rho(\abs{\xi_2},\abs{\xi_2}')=\int\limits_0^{\infty}d\phi \int\limits_{0}^{2\pi}d\abs{\xi_1}\,\Phi^*(\abs{\xi_1},\abs{\xi_2'},\phi)\Phi(\abs{\xi_1},\abs{\xi_2},\phi)
\end{equation}
The above integral cannot be solved analytically. We will therefore look at a special case where we only consider the correlations between points that are very close to each other. We may therefore write $\abs{\xi_2}^2/\abs{\xi_2}'^2=1-\epsilon$, where $\epsilon$ is very small. We also write the integral in terms of $\zeta_1=\abs{\xi_1}/\abs{\xi_2}$. The resulting density matrix will have the form:
\begin{equation}
    \rho(\abs{\xi_2},\abs{\xi_2'})=2I(\abs{\xi_2},\abs{\xi_2'})\exp{-\frac{1}{2}(\abs{\xi_2}^2+\abs{\xi_2'}^2)},
    \end{equation}
    where we will have to solve the integral $I(\abs{\xi_2},\abs{\xi_2'})$ given below:
\begin{multline}\label{bin}
   I(\abs{\xi_2},\abs{\xi_2'})=\abs{\xi_2}^{7/2}\abs{\xi_2'}^{3/2}\int d\zeta_1\,\zeta_1(1+\zeta_1^2)\\\times\bigg(1-\frac{\epsilon\zeta_1^2}{1+\zeta_1^2}\bigg)^{1/2}\exp{-\zeta_1^2\abs{\xi_2}^2}
\end{multline}
Since $\epsilon$ is very small, we may use binomial expansion to rewrite the integral in a solvable form as follows:
\begin{multline}
    I(\abs{\xi_2},\abs{\xi_2'})\approx\abs{\xi_2}^{7/2}\abs{\xi_2'}^{3/2}\int d\zeta_1\bigg(\zeta_1+\zeta_1^3-\frac{\epsilon\zeta_1^3}{2}\\-\frac{\epsilon^2\zeta_1^5}{8(1+\zeta_1^2)}-\frac{\epsilon^3\zeta_1^7}{16(1+\zeta_1^2)^2}\bigg)\exp{-\zeta_1^2\abs{\xi_2}^2}
\end{multline}
The reduced density matrix is now of the form:
%\begin{widetext}
\begin{multline}
    \rho(\abs{\xi_2},\abs{\xi_2'})=\frac{\abs{\xi_2'}^{3/2}}{16\abs{\xi_2}^{1/2}}\exp{-\frac{1}{2}(\abs{\xi_2}^2+\abs{\xi_2'}^2)} \\ \bigg(16-8\epsilon-2\epsilon^2-\epsilon^3+2(8+\epsilon^2+\epsilon^3)\abs{\xi_2}^2+\epsilon^3\abs{\xi_2}^4\\+\epsilon^2(2+3\epsilon+\epsilon\abs{\xi_2}^2)\abs{\xi_2}^4\ei{(-\abs{\xi_2}^2)}\exp{\abs{\xi_2}^2}\bigg).
\end{multline}
%\end{widetext}

From the resultant form of the reduced density matrix, we are able to split the function as $\rho(\abs{\xi_2},\abs{\xi_2'})=h(\abs{\xi_2})g(\abs{\xi_2'})$. In order to find the eigenvalues of $\rho(\abs{\xi_2},\abs{\xi_2'})$, we need to solve the following integral equation: 
\begin{equation}
    \int d\abs{\xi_2'}\rho(\abs{\xi_2},\abs{\xi_2'})f_n(\abs{\xi_2'})=\lambda_n f_n(\abs{\xi_2})
\end{equation}
We may guess the form of $f_n(\abs{\xi_2})$ here as $h(\abs{\xi_2})$. The resultant eigenvalue is therefore given by:
\begin{align}
    \lambda&=\int\limits_{0}^{\infty}d\abs{\xi_2}\,h(\abs{\xi_2})g(\abs{\xi_2})\nonumber\\
    &=1-\frac{\epsilon}{4}-\frac{\epsilon^2}{24}-\frac{\epsilon^3}{64}
\end{align}
The entropy can now be calculated from the equation $S=-\lambda\ln{\lambda}$, which results in:
\begin{equation}
    S=-\bigg(1-\frac{\epsilon}{4}-\frac{\epsilon^2}{24}-\frac{\epsilon^3}{64}\bigg)\ln{\bigg\{1-\frac{\epsilon}{4}-\frac{\epsilon^2}{24}-\frac{\epsilon^3}{64}\bigg\}}
\end{equation}

\begin{figure}[ht]
    \centering
    \includegraphics[scale=0.6]{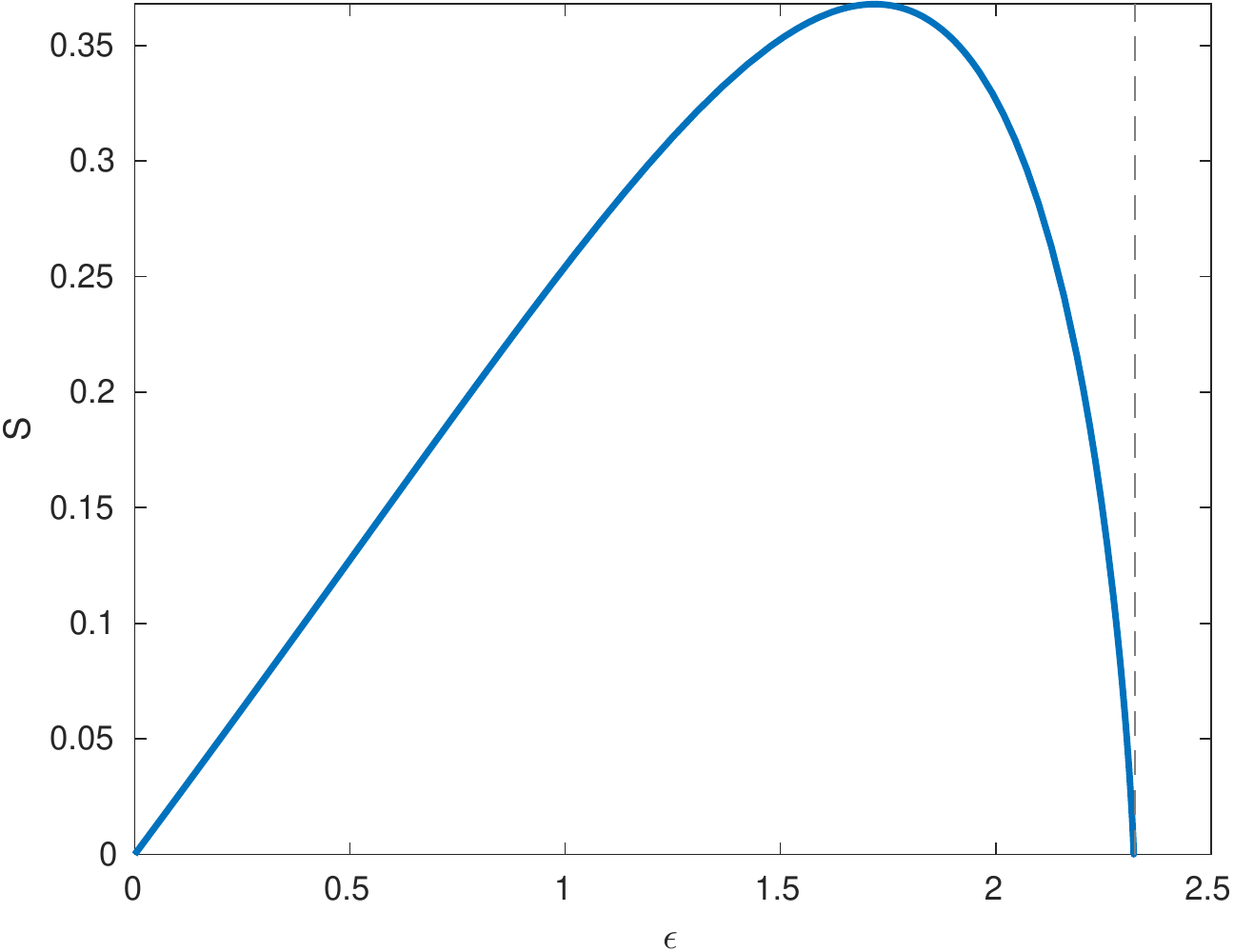}
    \caption{Plot of entanglement entropy in oscillator coordinates with respect to $\epsilon$, in the case where only correlations between closely spaced points are considered.}
    \label{fig4}
\end{figure}

The above results tell us that when we consider the correlations between very closely spaced points (very small values of $\epsilon$), the mapped system appears to be entangled since ($\abs{\xi_1},\abs{\xi_2},\phi$) is a distorted coordinate system with respect to the Hydrogen atom. If we take the case $\epsilon=0$, then the reduced density matrix is diagonal, resulting in zero entropy. But closer to zero, entropy varies linearly with respect to $\epsilon$, as can be seen from \ref{fig4}. However, information about the divergence of entropy in the ($\abs{\xi_1},\abs{\xi_2},\phi$) coordinate system can only be obtained by solving (\ref{red}) analytically, while also accounting for the dimensional parameters.

\section{Isotonic Oscillator Mapping}
\label{app:b}

The Hamiltonian of the quantum isotonic oscillator is  \cite{0305-4470-20-13-034}
 \begin{equation}
       H=\frac{p^2}{2m}+\frac{1}{2}m\tilde{\omega}^2x^2+\frac{g}{2mx^2}
   \end{equation}
 It should be noted that this Hamiltonian is similar to that of a spherical harmonic oscillator, and we may define $g=\hbar^2l(l+1)$ for convenience, where $l$ is a non-negative real number. The energy eigenvalues of the oscillator are given by $E_n=\hbar\tilde{\omega}(2n+l+3/2)$, and ground state wave-function is given by $\chi_0(x)=Cx^{l+1}\exp{-\beta x^2/2}$, where $\beta=m\tilde{\omega}/\hbar$. Under a certain set of transformations, we can map the radial part of Coulomb sub-system to an isotonic oscillator. For the ground state, $\phi_{int}(\vec{r})=R(r)/\sqrt{4\pi}$, where $R(r)$ can be found by solving the following equation for $u(r)$:
 \begin{equation}\label{app1}
    \bigg\{-\frac{\hbar^2}{2m}\frac{\partial^2}{\partial r^2}-\frac{e^2}{r}\bigg\}u(r)=-Bu(r),
\end{equation}
where $u(r)=rR(r)$. Now let us make the transformations $r=x^2$ and $u(x)=\sqrt{x}\chi(x)$. The latter requires that we consider only the non-negative half of the $x$-axis. On further rearranging ($\ref{app1}$), we get:
   \begin{equation}
       \bigg\{-\frac{\hbar^2}{2m}\frac{\partial^2}{\partial x^2}+4B x^2+\frac{3\hbar^2}{8mx^2}\bigg\}\chi(x)=4e^2\chi(x)
   \end{equation}
Comparing with the Hamiltonian for a 1-D isotonic oscillator, we can see that $E_n=4e^2$, $\tilde{\omega}=\sqrt{8B/m}$, and $l=1/2$. We can also rearrange the energy eigenvalue equation to arrive at the Rydberg formula $B=e^4m/2\hbar^2(n+1)^2$. From this it is clear that the ground state of Hydrogen atom corresponds to the ground state ($n=0$) of the isotonic oscillator, and the wave-function is given by $\chi(x)=Cx^{3/2}e^{-\beta x^2/2}$. The constant $C=\beta^{3/2}/2$ is found from the condition 2$\int_0^\infty dx\,x^2\abs{\chi(x)}^2=1$, which also preserves the Hydrogen atom normalization condition. The total wave-function for the Hydrogen atom in the original coordinates after the mapping has been done is given below:
\begin{multline}
    \Psi(\vec{r_e},\vec{r_p})=\frac{\beta^{3/2}}{\sqrt{8\pi\Omega}}\exp{-\frac{\beta}{2}\abs{\vec{r_e}-\vec{r_p}}}\\\times\exp{i\frac{\vec{P}}{\hbar}.\frac{m_e\vec{r_e}+m_p\vec{r_p}}{M}},
\end{multline}
where $\beta=m\tilde{\omega}/\hbar$. The reduced density matrix for the electronic sub-system is of the form:
\begin{equation}
\rho(\vec{r_e},\vec{r_e'})=\int d^3\vec{r_p}\Psi^*(\vec{r_e'},\vec{r_p})\Psi(\vec{r_e},\vec{r_p})
\end{equation}
On proceeding to find the eigenvalues of reduced density matrix as was done in Section \hyperref[5]{V}, we get:
\begin{equation}
    \tilde{\rho}(\vec{k})=\frac{1}{\Omega}\frac{64\pi a_0^3}{\big(1+a_0^2\abs{\vec{k}-\vec{k_e}}^2\big)^4}
\end{equation}
The eigenvalues obtained are exactly the same as in the unmapped case (\ref{eig1}). It can therefore be deduced that IR-divergence of entropy is preserved in this particular mapping.

\section{Massive scalar field (1+1) dimensions}\label{appc}
For N=3, the Hamiltonian in (\ref{1d}) becomes:
\begin{equation}
    H^{(\text{1D})}=\frac{1}{2}\sum_{i=0}^2\bigg\{\frac{\pi^2}{a}+\bigg(am_f^2+\frac{2}{a}\bigg)\phi_i^2-\sum_{j\neq i}\frac{1}{a}\phi_i\phi_j\bigg\}
\end{equation}
The potential matrix is:
\begin{equation}
    K=\begin{bmatrix}m_f+2/a^2&-1/a^2&0\\-1/a^2&m_f+2/a^2&-1/a^2\\0&-1/a^2&m_f+2/a^2\end{bmatrix}
\end{equation}
The normal modes are found to be:
\begin{align}
    \omega_0^2&=m_f^2+\frac{2}{a^2}\nonumber\\
    \omega_1^2&=m_f^2+\frac{2-\sqrt{2}}{a^2} \nonumber
    \end{align}
   \begin{align}  
    \omega_2^2&=m_f^2+\frac{2+\sqrt{2}}{a^2}
\end{align}
None of the above modes vanish unless we consider the case $m_f\to0$ and $a\to\infty$ simultaneously. Since there are no zero-modes in the massive case, entanglement entropy diverges in the UV limit as opposed to the IR limit.

\end{document}